\documentclass[12pt,a4paper,final]{iopart}
\usepackage{iopams}  
\usepackage{graphicx}
\usepackage{subfig}
\usepackage[titletoc,title]{appendix}
\usepackage[dotinlabels]{titletoc}
\usepackage[breaklinks=true,colorlinks=true,linkcolor=blue,urlcolor=blue,citecolor=blue]{hyperref}

\begin{document}
\title{Stochastic Efficiency: Five Case Studies}

\author{Karel Proesmans and Christian Van den Broeck}
\address{Hasselt University, B-3590 Diepenbeek, Belgium}
\ead{Karel.proesmans@uhasselt.be}

\begin{abstract}
Stochastic efficiency is evaluated in five case studies: driven Brownian motion, effusion with a thermo-chemical and thermo-velocity gradient, a quantum dot and a model for information to work conversion. The salient features of stochastic efficiency, including the maximum of the large deviation function at the reversible efficiency, are reproduced. The approach to and extrapolation into the asymptotic time regime are documented.
\end{abstract}

\vspace{2pc}
\submitto{\NJP}

The concept of Carnot efficiency is a founding principle of macroscopic thermodynamics. It allows to introduce entropy as a state function and to define the Kelvin temperature scale. It states that, for a system operating between two reservoirs at temperatures $T_h$ and $T_c$,  the efficiency $\bar{\eta}=W/Q_h$, being the ratio of output work $W$ over input heat $Q_h$, is bounded by the Carnot efficiency $\bar{\eta} \leq \eta_C$ with $\eta_C=1-T_c/T_h$.  Right from the start, the question was raised about the efficiency of small scale machines. Maxwell introduced a small-scale demon which was deemed to rectify thermal fluctuations.  A clarifying rebuttal was given by Smoluchowski, who proposed a mechanical implementation of the Maxwell demon, the so-called ratchet and pawl. He stressed that this small-scale device would eventually thermalize, after which any rectification would stop. Szilard introduced an information driven engine which seemed to be able to extract work from a single reservoir, in violation with the Carnot prediction. Since he believed that this could not be true, he concluded that there had to be a thermodynamic cost associated to the information gathering \cite{szilard1929entropieverminderung}.  The Szilard engine gave rise to a prolonged scientific discussion about the source of dissipation \cite{leff2002maxwell}.  It is not the measurement  or computational process, but the resetting process of the memory device that seems to be the dissipative step. The Smoluchowski engine was revisited by Feynman, who showed by an explicit model calculation that the efficiency of the ratchet and pawl in contact with two reservoirs is indeed bounded by the Carnot efficiency. A subtle error in his analysis was elucidated by Parrondo and Sekimoto \cite{parrondo1996criticism,sekimoto1997kinetic}, indicating that the efficiency was strictly below Carnot efficiency in this model. This was confirmed by a more detailed analysis on simplified models \cite{jarzynski1999feynman,van2004microscopic}, prompting the question whether Carnot efficiency could at all be reached in such small devices \cite{sekimoto1998langevin}. These questions have led to major interest in information to work conversion, both theoretically \cite{barato2014unifying,barato2014stochastic,mandal2012work,touchette2000information,sagawa2010generalized,horowitz2011designing,kish2012energy,abreu2012thermodynamics,mandal2013maxwell,esposito2012stochastic,horowitz2013imitating} and experimentally \cite{raizen2009comprehensive,raizen2011demons,berut2012experimental,thorn2008experimental,Toyabe,strasberg2013thermodynamics,Serreli}.

Thermodynamic efficiency can also be defined for other types of engines, notably  for work to work and the above mentioned information to work transformations. In the case of a transformation of input work $W_i$ to output work $W_o$, the thermodynamic efficiency is defined as $\bar{\eta}=W_o/W_i$. The analogue of Carnot efficiency is reached for a reversible operation leading to $W_o=W_i$ (since no heat is dissipated). Hence the second law stipulates $\bar{\eta}\leq{\eta}_{r}$, with the reversible efficiency is ${\eta}_{r}=1$. For information to work transformation, we note that information about the system can be used to extract work. The reversible limit has been known since the work of Szilard, namely $k_B T \ln 2$  of work can be extracted per bit of information (in an environment operating at temperature $T$). One bit corresponds to a Shannon information of $I=\ln2$. More generally, in a transformation of  a Shannon information amount $I$ into an amount $W$ of work, the efficiency  $\bar{\eta}=W/(k_B T I)$ is upper bounded by the reversible result ${\eta}_{r}=1$.

Over the past two decades, one has been able to reformulate thermodynamics to describe fluctuating small-scale systems. The most notable of the results is the fluctuation theorem, stating that the probability for a stochastic entropy production $\Delta_is$ is exponentially more probable than that of  a corresponding decrease $-\Delta_is$ in an "inverse" experiment:  
\begin{equation}\label{ft}
\frac{P(\Delta_is)}{\tilde{P}(-\Delta_is)}=\exp(k_B \Delta_is).
\end{equation}
The tilde refers to the ``time- inverse" experiment. 
The above symmetry relation for the probability implies the following integral fluctuation theorem:
\begin{equation}\label{ift}
\langle \exp{(-k_B \Delta_is)}\rangle=1,
\end{equation}
which in term implies the ``second law":
\begin{equation}
\langle\Delta_i s \rangle \geq 0,
\end{equation}
The detailed and integral fluctuation theorems  (\ref{ft}) and  (\ref{ift}) have been derived in many different settings, see for example the Jarzynski \cite{jarzynski1997nonequilibrium} and Crooks \cite{crooks1999entropy} equalities, and stochastic thermodynamics \cite{seifert2012stochastic,vandenbroeck}. When considering a finite time experiment, the interpretation of  (\ref{ft}) is somewhat delicate, and there may be restrictions, for example on the initial and final states of the experiment, cf.~\cite{becker2015echo} for a recent discussion. The fluctuation theorem however appears to have a wide ranging validity, comparable to the second law, when considering the asymptotic long time limit, in which case (\ref{ft}) reduces to a statement about the large deviation properties of the entropy production. The theorem was actually first derived in this context \cite{evans1993probability,gallavotti1995dynamical,lebowitz1999gallavotti,kurchan1998fluctuation}. Surprisingly, the impact of these new insights on  the efficiency, and in particular on its stochastic properties, has only been considered very recently \cite{nc,berkeley,verley2014universal,proesmans2014stochastic,polettini2014finite,rana2014single,campisi2014nonequilibrium,esposito2015efficiency,martinez2014brownian}, and only in work to work and heat to work converters. When running a  small-scale engine for a finite time, the corresponding cumulated work output and heat  uptake, $w$ and $q_h$, or work output $w_o$ and work input $w_i$, are stochastic quantities, and hence so is the corresponding stochastic efficiency ${\eta}={w}/q_h$ or ${\eta}={w_o}/w_i$. 
%The work and heat converge in the sense of the law of large numbers to their average values $W=\bar{w}$ and $Q_h=\bar{q_h}$ in the limit of a large ensemble or long time average. As the ensemble average describes a macroscopic machine, the corresponding efficiency is upper bounded by Carnot efficiency: $\bar{\eta}={W}/Q_h\leq{\eta}_C$.
Starting with the detailed fluctuation theorem (\ref{ft}), it was pointed out that one can make universal statements about the stochastic efficiency ${\eta}$, as it approaches the macroscopic efficiency. More precisely, for large but finite times, values of the stochastic efficiency $\eta$ different from the macroscopic  efficiency $\bar{\eta}$ are exponentially unlikely, as described by the large deviation function $J(\eta)=-\lim_{t\rightarrow\infty}1/t\ln P_t(\eta)$, $J(\eta)\geq J(\bar{\eta})=0$. For the case of a thermodynamic machine driven by a time-symmetric protocol, it was shown that the reversible efficiency is the least likely, i.e., the large deviation function  $J(\eta)$ has a maximum at $\eta=\eta_{r}$. The macroscopic efficiency is reproduced as the value carrying all probability to dominant order  $J(\bar{\eta})=0$. For time-asymmetric protocols, the large deviation curves of the stochastic efficiency for the forward and backward experiment cross at the reversible efficiency \cite{berkeley,verley2014universal}. 

In this paper, we provide an explicit and comprehensive illustration of the stochastic efficiency in five engines. Providing possibly the  simplest steady state models for work to work and heat to work transformation, we consider a Brownian particle subject to competing forces \cite{nc} and effusion between two compartments \cite{proesmans2014stochastic}. Even though these cases have been discussed in some detail in the literature, they are introduced briefly for completeness and for comparison with the other models. The third model is a thermal engine based on a quantum dot. It is of interest because its stochastic thermodynamic properties have been discussed and measured, and because it is in principle richer than the effusion model, which can be recovered in an appropriate mathematical limit. In the last two problems, we evaluate stochastic efficiency in novel settings. We show that the universal features of stochastic efficiency are valid for heat to momentum transformation.  This is illustrated on an effusion model with momentum transfer. Finally, we discuss the stochastic efficiency for information to work transformation. It is indeed possible to reformulate the fluctuation theorem when dealing with an information processing set-up, such as the one introduced by Szilard. In particular one can introduce the stochastic efficiency $\eta=w/i$ for a machine transforming a stochastic amount of input (Shannon) information $i$ into work $w$. We show that its large deviation function again displays the same general features. In particular, the reversible efficiency $\eta_{r}=1$ is exponentially less likely than any other efficiency in the asymptotic time limit (for time-symmetric protocols).  We illustrate these features on the Mandal-Jarzynski model \cite{barato2014unifying,mandal2012work}.

\section{Brownian Engine}
\begin{figure}\begin{centering}
\subfloat[Schematic representation of a Brownian work-to-work converter]{\includegraphics[width=6cm]{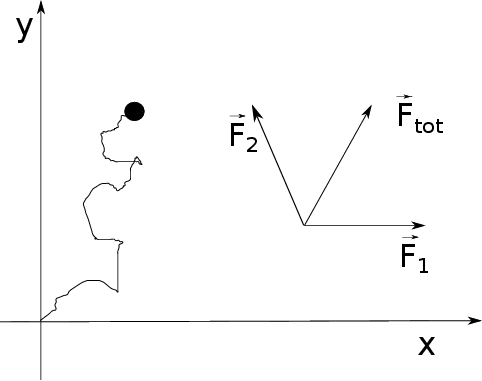}\label{fig1}}
\qquad
\subfloat[Colour-coded macroscopic efficiency of a Brownian engine in function of $\vec{F}_2$ with $\vec{F}_1=(1,0)$. The green dot corresponds to $\vec{F}_2=(-3/2,1)$ and   $\bar{\eta}=0.29$. The stochastic efficiency for this case is represented in Fig.~\ref{fig3}.]{\includegraphics[width=6cm]{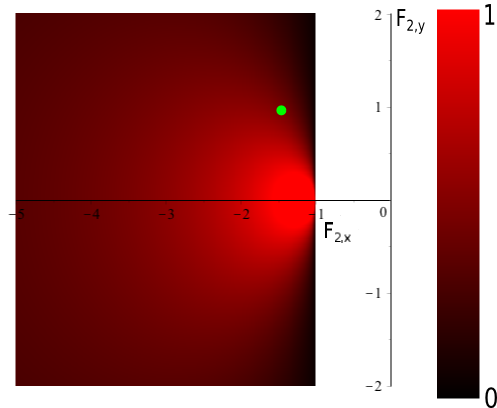}\label{fig2}}
\caption{Characteristics of the Brownian engine}
\end{centering}\end{figure}

Consider an overdamped Brownian particle on a plane, subject to two external forces, a loading force $\vec{F_1}$, and a driving force $\vec{F_2}$ \cite{nc}, cf.~Fig.~\ref{fig1}. The (larger) driving  force $\vec{F_2}$ pushes the particle against the loading force $\vec{F_1}$. The stochastic efficiency of such a device  as a work-to-work converter  was discussed in \cite{nc,polettini2014finite}.
The mathematics are very simple. Considering for simplicity a two dimensional set-up, the displacement $\vec{x}$ of the Brownian particle during a time $t$ is characterised by a bi-Gaussian distribution.  Under influence of the resulting force $\vec{F}=\vec{F_1}+\vec{F_2}$  the average displacement is $\left\langle \vec{x}\right\rangle=\mu \vec{F} t$,  $\mu$ being the mobility. The variance is isotropic and uncorrelated in orthogonal directions, $\left\langle\delta\vec{x}\delta\vec{x}\right\rangle=2Dt \vec{1}$, where $D$ is the diffusion coefficient and $\vec{1}$ the unit matrix. 

We first turn to the macroscopic efficiency, which is very easy to evaluate, see also Fig.~\ref{fig2} for a colour-coded illustration:
\begin{equation}\label{ee}\bar{\eta}=-\frac{\left\langle w_1\right\rangle}{\left\langle w_2\right\rangle}=-\frac{\vec{F}_1\cdot\left\langle \vec{x}\right\rangle}{\vec{F}_2\cdot\left\langle  \vec{x}\right\rangle}=-\frac{\vec{F}_1\cdot\vec{F}}{\vec{F}_2\cdot\vec{F}},\end{equation}
where $w_1=\vec{F}_1\cdot \vec{x}$ and $w_2=\vec{F}_2\cdot \vec{x}$ are the stochastic amounts of work delivered by the loading and driving force respectively.
The engine regime, i.e., the regime where the driving force delivers a positive amount of work to the loading force, is determined by:
\begin{equation}\label{EngineCondBrown}-\frac{\left|\vec{F_2}\right|}{\left|\vec{F}_1\right|}\leq \cos\theta\leq -\frac{\left|\vec{F}_1\right|}{\left|\vec{F}_2\right|},\end{equation}
with $\theta$ is the angle between $\vec{F}_1$ and $\vec{F}_2$. In combination with (\ref{ee}), it is clear from (\ref{EngineCondBrown}) that, in the engine regime, the macroscopic efficiency is bounded by $\bar{\eta}\leq \eta_r=1$. The reversible efficiency $\eta_r=1$ can only be reached in the limit $\vec{F}_2\rightarrow -\vec{F}_1$ with $\vec{F}_2 \parallel -\vec{F}_1$. This can also be seen in Fig.~\ref{fig2}.

We next investigate the stochastic efficiency $\eta=-w_1/w_2=-\vec{F}_1\cdot\vec{x}/\vec{F}_2\cdot\vec{x}$. Being the ratio of two correlated Gaussian variables, its probability distribution can be evaluated analytically, see also  \cite{polettini2014finite}:
\begin{equation}P_t(\eta)=\frac{\left|\vec{F}_1\times \vec{F}_2\right| e^{-\frac{t}{t_0}}}{(\vec{F}_{1}+\eta \vec{F}_{2})^2\pi}\left(1+\sqrt{\pi g(\eta)} \mathrm{Erf}(\sqrt{g(\eta)})e^{g(\eta)}\right),\label{PetaGauss}\end{equation}
with
%\begin{equation}f(\eta)=\frac{\left|\vec{F}_1\times \vec{F}_2\right|}{(\vec{F}_{1}+\eta \vec{F}_{2})^2\pi},\end{equation}
\begin{equation}g(\eta)=\frac{t}{t_0}\frac{\left(1-\eta\right)^2\left(\vec{F_1}\times \vec{F}_2\right)^2}{\vec{F}^2\left(\vec{F}_1+\eta\vec{F}_2\right)^2},\label{GetaGauss}\end{equation}
and $\mathrm{Erf(x)}$ is the error function. The characteristic time  $t_0={2D}/\left(\mu\left|\vec{F}\right|\right)^2$ determines the boundary between diffusion dominated ($t\ll t_0$) and drift dominated ($t\gg t_0$) dynamics. We note in passing that it is easy to show from equations (\ref{PetaGauss}) and (\ref{GetaGauss}) that $P'_t(0)>0$ and $P'_t(\eta_r)<0$, implying that there exists at least one maximum in the interval $\eta\in\left[0,1\right]$
  
\begin{figure}\begin{centering}
\subfloat[Probability distribution $P_t(\eta$) of the efficiency $\eta$ for the Brownian engine.]{\includegraphics[width=6cm]{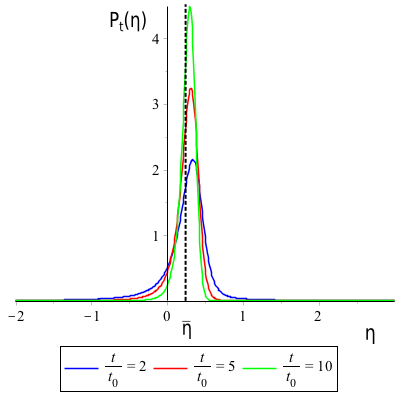}\label{fig3}}
\qquad
\subfloat[Approach of $-\ln P_t(\eta)  /t$  for increasing values of $t/t_0$ (blue, red and green curve)  to the large deviation function $J(\eta)$ (black curve). The yellow curve is obtained by extrapolation from the finite time results.]{\includegraphics[width=6cm]{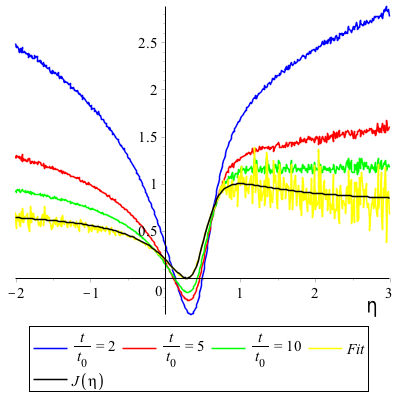}\label{fig4}}
\caption{Efficiency fluctuations of a Brownian engine with $\vec{F}_1=(1,0)$, $\vec{F}_2=(-3/2,1)$ and $\bar{\eta}=0.29$.}
\end{centering}\end{figure} 
%These results indeed seem to agree with simulation results, cf.~Fig.~\ref{fig3}. 

Universal features of the efficiency fluctuations are revealed when studying  the asymptotic time behavior via the large deviation function of $\eta$  \cite{nc}:
\begin{eqnarray}\label{LDFBrown}J(\eta)&=&-\lim_{t\rightarrow\infty}\frac{1}{t}\ln P_t(\eta)\nonumber\\&=&\frac{\mu^2}{4D}\frac{\left[(\vec{F_1}+\eta\vec{F_2})\cdot(\vec{F_1}+\vec{F_2})\right]^2}{(\vec{F_1}+\eta\vec{F_2})^2}.\end{eqnarray}
%due to the Cauchy-Schwarz inequality. 
This function has a minimum at the macroscopic efficiency $J(\bar{\eta})=0$, but also a maximum at the reversible efficiency $\eta_r=1$, with equal asymptotes in the limits $\eta\rightarrow\pm\infty$.
To illustrate the approach to the large deviation regime, $-\ln(P_t(\eta))/t$ is plotted in Fig.~\ref{fig4} for $t/t_0=2, 5,$ and $10$, together with the limiting expression (\ref{LDFBrown}). We also include the result of an extrapolation ansatz \cite{proesmans2014stochastic}, described in more detail in \ref{AppLDF}. The extrapolation, based on the $t/t_0=2, 5,$ and $10$ curves, is in surprisingly good agreement with the exact asymptotic expression. Although these finite time results do not, in this particular instance, exhibit a maximum close to the reversible efficiency $\eta_r=1$, it does show up by extrapolation.

\section{Effusion Engine}
\begin{figure}\begin{centering}
\subfloat[Schematic representation of the effusion model.]{\includegraphics[width=6cm]{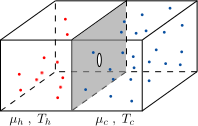}\label{fig5}}
\qquad
\subfloat[Macroscopic efficiency of the effusion engine in terms of $\mu_h/k_BT_h$ and $\mu_c/k_B T_h$, with $\eta_C=1/2$. The green dot corresponds to $\mu_h/k_BT_h=-1$ and $\mu_c/k_BT_h=-3/4$. The stochastic efficiency for this case is represented in Fig.~\ref{fig8}.]{\includegraphics[width=6cm]{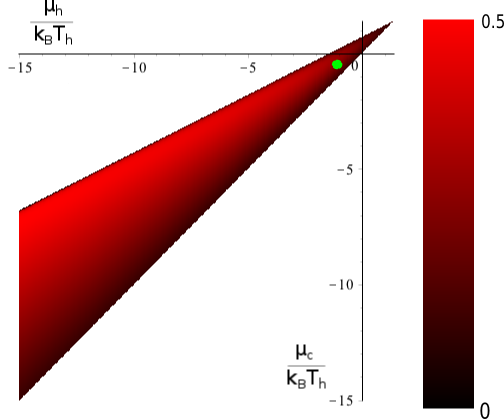}\label{fig6}}
\caption{Characteristics of the effusion model}
\end{centering}\end{figure}
The effusion engine \cite{proesmans2014stochastic} consists of two reservoirs, exchanging heat and particles by effusion via one or more small holes in the separating wall, cf.~Fig.~\ref{fig5}. The reservoirs are supposed to be infinitely large and at equilibrium. The holes are smaller than the mean free path so that the equilibrium state is not disturbed by the effusion process.  Under proper working conditions, a net flux of particles moves  from say the left compartment, at high temperature $T_h$ and low chemical potential  $\mu_h$, to the right compartment  at lower temperature $T_c$ but higher chemical potential $\mu_c$.  

When a particle moves from the hot to the cold reservoir, it delivers an amount of work $w_0=\mu_c-\mu_h=\Delta \mu$, while extracting an amount of heat $q_0=u_0-\mu_h$ from the hot reservoir,
where $u_0$ is the kinetic energy of the transfered particle. Therefore, after a net transfer of $n$ particles, the total amount of delivered work and extracted heat, $w$ and $q$ are given by:
\begin{eqnarray}w=n\Delta \mu,\end{eqnarray}
\begin{eqnarray}q=u- n\mu_h,\end{eqnarray}
with $u$ the net energy transfers.

Particles with kinetic energy $E$ transfer from the hot to the cold reservoir at rate \cite{preCleuren}:
\begin{equation}\label{EffThc}T_{h\rightarrow c}(E)=\frac{1}{t_0}\frac{E}{\left(k_B T_h\right)^2}e^{-\frac{E}{k_B T_h}},\end{equation}
and from the cold to the hot reservoir at rate
\begin{equation}\label{EffTch}T_{c\rightarrow h}(E)=\frac{1}{t_0}\frac{E}{\left(k_B T_h\right)^2}e^{-\frac{E}{k_B T_c}+\frac{\mu_c}{k_BT_c}-\frac{\mu_h}{k_B T_h}},\end{equation}
where $t_0=\sqrt{2\pi m/\left(\sigma^2\rho^2_hk_BT_h\right)}$ is the average time between particle crossings from the hot to the cold reservoir, $\sigma$ is the surface area of the effusion hole and $m$ is the mass of the particles of the gas.
The macroscopic efficiency can be easily obtained:
\begin{eqnarray}\bar{\eta}&=&\frac{\left\langle w\right\rangle}{\left\langle q\right\rangle}\nonumber\\&=&\frac{\Delta \mu\left\langle n \right\rangle}{\left\langle u\right\rangle-\mu_h \left\langle n\right\rangle }\nonumber\\&=&\frac{\Delta \mu \left((k_BT_h)^2 e^{\frac{\mu_h}{k_BT_h}}-(k_BT_c)^2 e^{\frac{\mu_c}{k_BT_c}}\right)}{(k_BT_h)^2  (2 k_BT_h-\mu_h)e^{\frac{\mu_h}{k_BT_h}}-(k_BT_c)^2\left(2 k_BT_c-\mu_h \right)e^{\frac{\mu_c}{k_BT_c}} }.\end{eqnarray}
This is plotted for Carnot efficiency $\eta_C=1/2$ in Fig.~\ref{fig6} in terms of $\mu_c/k_B T_h$ and $\mu_h/k_BT_h$. The engine boundaries are given by $\mu_h<\mu_c<\mu_hT_c/T_h-2k_B T_c \ln\left(T_c/T_h\right)$. Furthermore, the macroscopic efficiency is bounded by the Carnot efficiency, and this boundary is only reached in the limit $\mu_c,\mu_h\rightarrow -\infty$, which, for an ideal gas, corresponds to zero density.

\begin{figure}\begin{centering}
\subfloat[Probability distribution $P_t(\eta$) of the efficiency $\eta$ for the effusion engine.]{\includegraphics[width=6cm]{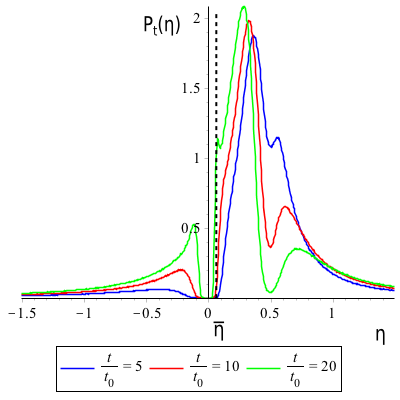}\label{fig7}}
\qquad
\subfloat[Approach of $-\ln P_t(\eta)  /t$  for increasing values of $t/t_0$ (blue, red and green curve)  to the large deviation function $J(\eta)$ (black curve). The yellow curve is obtained by extrapolating the finite time results.]{\includegraphics[width=6cm]{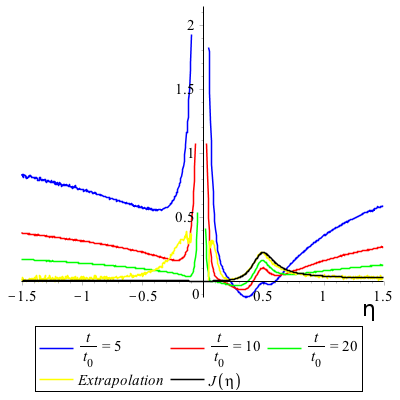}\label{fig8}}
\caption{Efficiency fluctuations of the effusion engine, with $\eta_C=1/2$, $\mu_h=-k_B T_h$, and $\mu_c=-3/4\,k_B T_h$. The macroscopic efficiency is given by $\bar{\eta}=0.07$.}
\end{centering}\end{figure}

We  next investigate the stochastic efficiency $\eta=w/q$. Since it is not possible to obtain the analytic expression of  the probability distribution $P_t(\eta)$, we  present results from a numerical simulation of the Markov process with the prescribed rates (\ref{EffThc}) and (\ref{EffTch}), cf.~Fig.~\ref{fig7}. 
For the parameter values under consideration, one clearly sees a minimum in the probability distribution developing in the vicinity of the Carnot efficiency $\eta_C$, even at the rather short times represented here. The behavior around $\eta=0$, and in particular the other minimum around $\eta=0$,  can be explained by the low number of particle crossings for short times \cite{proesmans2014stochastic}.
%: due to the fact that only a few particles will cross, small, non-zero efficiencies can only be reached if particles cross with very high energies. As this is very unlikely, the finite time probability distribution function show a minimum around zero. This effect will disappear in the large time limit. 

Turning finally to the asymptotic time behaviour, we note that the large deviation function $J(\eta)$ of the stochastic efficiency can be obtained from the joint cumulant generating function of work and heat $\varphi(\lambda,\omega)$, cf.~\cite{verley2014universal}. The latter is explicitly known for effusion \cite{preCleuren} :
\begin{eqnarray}
\varphi(\lambda,\omega)&=&\lim_{t\rightarrow\infty}\frac{1}{t}\ln\left\langle e^{-\lambda W-\omega Q}\right\rangle\nonumber\\&=&-\frac{\sigma \rho_h \sqrt{k_BT_h}}{\sqrt{2\pi m}}\left(1-\frac{\exp\left(-\lambda \Delta \mu - \omega \mu_h\right)}{(1-k_BT_h \omega)^2}\right)\nonumber\\&&-\frac{\sigma \rho_c \sqrt{k_BT_c}}{\sqrt{2\pi m}}\left(1-\frac{\exp\left(\lambda \Delta \mu + \omega \mu_c\right)}{(1+k_BT_c \omega)^2} \right).
\end{eqnarray}
The large deviation function of the stochastic efficiency is then given by:
\begin{equation}
J(\eta)=-\min\limits_{\lambda}\varphi(\lambda,\lambda \eta).\label{JEff}
\end{equation}
The contraction can be done numerically, and the comparison with finite-time simulations is represented in Fig.~\ref{fig8}. Note  that the extrapolation again works quite well, except in the vicinity of $\eta=0$.

\section{Quantum Dot}
\begin{figure}\begin{centering}
\subfloat[Schematic represetation of the quantum dot.]{\includegraphics[width=6cm]{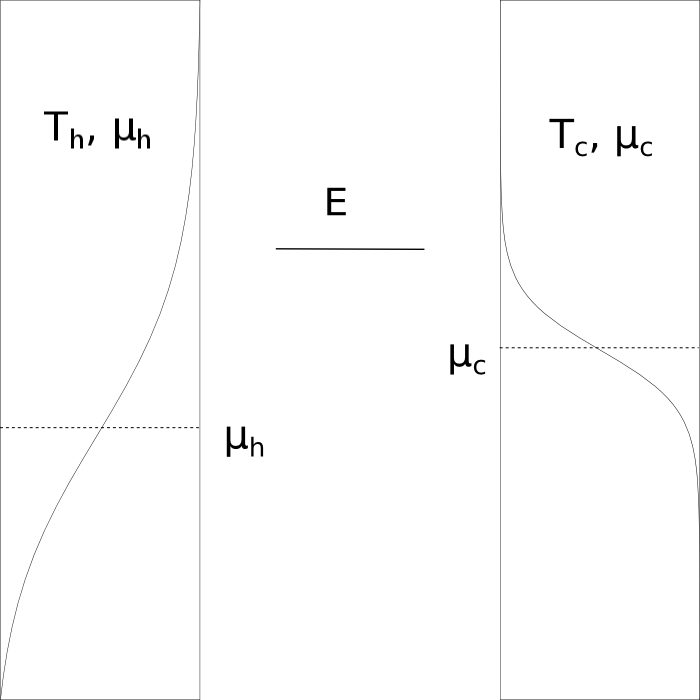}\label{fig9}}
\qquad
\subfloat[Macroscopic efficiency of the quantum dot, with $\eta_C=4/5$, $E_1=k_B T_h$ and $E_2=2\, k_B T_h$.]{\includegraphics[width=6cm]{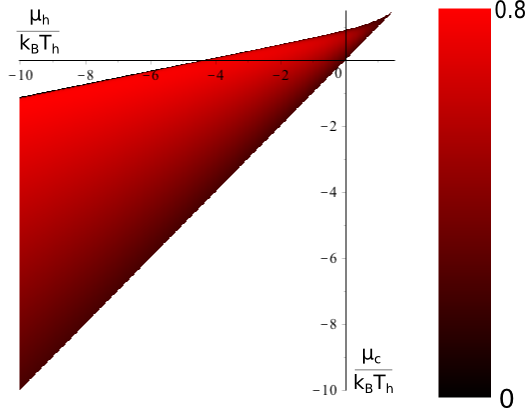}\label{fig10}}
\caption{Characteristics of the quantum dot}
\end{centering}\end{figure}

The quantum dot model,  schematically represented in  Fig.~\ref{fig9},  has been investigated in detail in the context of stochastic thermodynamics \cite{esposito2009thermoelectric,esposito2010finite,esposito2010quantum,esposito2012stochastically}. Two electron reservoirs are brought in contact with each other via one or multiple quantum dots. In order to investigate its stochastic efficiency, we focus on the case of two quantum dots, each with one ``active" energy level, $E_1$ and $E_2$ ($E_1<E_2$). Occupancy of a quantum dot by multiple electrons is forbidden because of Coulomb repulsion. For mathematical simplicity, we also set all coupling constants between dot and  reservoirs equal to $\Gamma$.

The operation of each quantum dot as a thermal engine is similar to that of the effusion engine: a net motion of electrons from  a reservoir with  low chemical potential   to one with higher chemical potential is induced by  a driving  temperature gradient. 
For every particle transferring through the quantum dot with the lower energy level $E_1$, the heat taken from the hot reservoir is given by $\delta q_1=E_1-\mu_h$ and the delivered amount of work is $w_0=\mu_c-\mu_h=\Delta\mu$. 
Furthermore, the rate of transfer between the hot reservoir and the quantum dot are given by:
\begin{equation}\label{RateQD1}k^{+}_h=\frac{\Gamma e^{-\frac{E_1-\mu_h}{T_h}}}{1+e^{-\frac{E_1-\mu_h}{T_h}}},\end{equation}
\begin{equation}\label{RateQD2}k^{-}_h=\frac{\Gamma}{1+e^{-\frac{E_1-\mu_h}{T_h}}}.\end{equation}
The rate of exchange between the cold reservoir and the quantum dot is obtained by replacing with $T_h$ and $\mu_h$  by $T_c$ and $\mu_c$, respectively. 
Analogous expressions hold for transfer through the other quantum dot, with $E_1$ replaced by $E_2$. The total amount of delivered work $w$ and consumed heat $q$ after a net transfer of $n_1$ particles through the quantum dot with energy level $E_1$ and $n_2$ particles through the quantum dot with energy level $E_2$ are then given by:
\begin{equation}w=\Delta \mu\left(n_1+n_2\right),\end{equation}
\begin{equation}q=\delta q_1 n_1+\delta q_2 n_2. \end{equation} 
The macroscopic efficiency can now be written as:
\begin{eqnarray}\bar{\eta}&=&\frac{\left\langle w\right\rangle}{\left\langle q\right\rangle}\\&=&\frac{\Delta\mu\left(\left\langle n_1\right\rangle+\left\langle n_2\right\rangle\right)}{\left\langle n_1 \right\rangle(E_1-\mu_c)+\left\langle n_2\right\rangle (E_2-\mu_c)},\end{eqnarray}
where $ n_1$ and $ n_2$ are the (stochastic) net amount of particles transferred through respectively the quantum dot with the lower and higher energy level. Their average value  can  be calculated from equations (\ref{RateQD1}) and (\ref{RateQD2}):
\begin{equation}\left\langle n_1\right\rangle=\frac{\Gamma}{2}\frac{{\rm e}^{{\frac {E_{{1}}-\mu_{{c}}}{T_{{c}}}}}-{
{\rm e}^{{\frac {E_{{1}}-\mu_{{h}}}{T_{{h}}}}}}}{\left( {
{\rm e}^{{\frac {E_{{1}}-\mu_{{c}}}{T_{{c}}}}}}+1 \right) \left( {{\rm e}^{{\frac {E_{{1}}-\mu_{{h}}}{T_{{h}}}}}}+1 \right) },\end{equation}
and an analogous expression for $\left\langle n_2\right\rangle$ with $E_1$ replaced by $E_2$.
The macroscopic efficiency of the engine is plotted in Fig.~\ref{fig10} in function of $\mu_h/k_BT_h$ and $\mu_c/k_BT_h$. These results are comparable with the results of the effusion engine. Again, Carnot efficiency is only reached in the limit of zero density.

\begin{figure}\begin{centering}
\subfloat[Probability distribution $P_t(\eta$) of the efficiency $\eta$ for the quantum dot engine.]{\includegraphics[width=6cm]{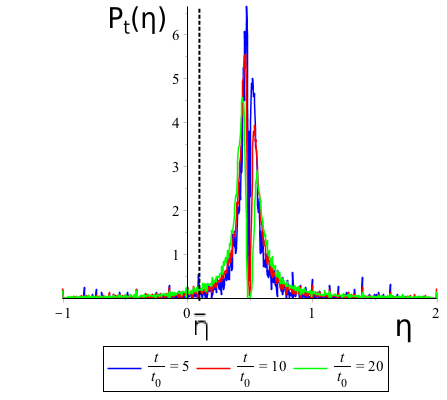}\label{fig11}}
\qquad
\subfloat[Approach of $-\ln P_t(\eta)  /t$  for increasing values of $t/t_0$ (blue, red and green curve)  to the large deviation function $J(\eta)$ (black curve). The yellow curve is obtained by extrapolation from the finite time results.]{\includegraphics[width=6cm]{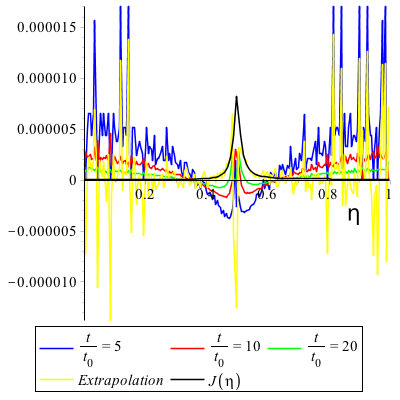}\label{fig12}}
\caption{Efficiency fluctuations of the quantum dot, with $\eta_C=1/2$, $E_1=k_B T_h$, $E_2=10\, k_B T_h$, $\mu_h=-k_BT_h$, $\mu_c=0$ and $t_0=10^5/\Gamma$. The macroscopic efficiency is given by $\bar{\eta}=0.09$.}
\end{centering}\end{figure}

For the study of the efficiency fluctuations at finite time, we again first turn to numerical simulations. 
The probability distribution $P_t(\eta)$,  $\eta=w/q$, is obtained by sampling the net fluxes $n_1$ and $n_2$ using the rates specified in equations (\ref{RateQD1}) and (\ref{RateQD2}). A typical result is shown in Fig.~\ref{fig11}. The results  appear to be rather noisy, which is due to the fact that the delivered amounts of work and heat are discrete variables. Nevertheless, the minimum at Carnot efficiency is very striking, even at these short times. The convergence to the macroscopic efficiency on the other hand is rather slow, which is, in this particular case, due to the small value of the large deviation function.

Due to the fact that this engine is driven by  two independent, tight-coupled operating channels (namely the two quantum dots), with heat consumption and efficiencies per particle $\delta q_i$ and $\eta_i=w_0/\delta q_i$ respectively, $i=1,2$, the large deviation function for the efficiency $J(\eta)$ can be written in terms of the event large deviation functions $\phi_i(n)$, $i=1,2$ (see \ref{AppEve}): 
\begin{equation}J(\eta)=\min_x\left(\phi_1\left(\frac{\alpha x}{\delta q_1}\right)+\phi_2\left(\frac{(1-\alpha)x}{\delta q_2}\right)\right).\label{EtaContrQD}\end{equation}
Here, $\alpha=(\eta-\eta_2)/(\eta_1-\eta_2)$, and $\phi_1 (n)$ and $\phi_2 (n)$ are the large deviation functions of the net number of transferred particles through each of the channels.
As these large deviation functions are known (see \ref{AppQD}), equation (\ref{EtaContrQD}) can be used to estimate the large deviation function of $\eta$. The comparison with finite time numerical simulations is shown in Fig.~\ref{fig12}. Note that the extrapolation from finite time results does not work uniformly well due to the aforementioned discreteness of work and heat variables.

\section{Effusion with momentum transfer}
\begin{figure}\begin{centering}
\subfloat[Schematic representation of effusion with momentum]{\includegraphics[width=6cm]{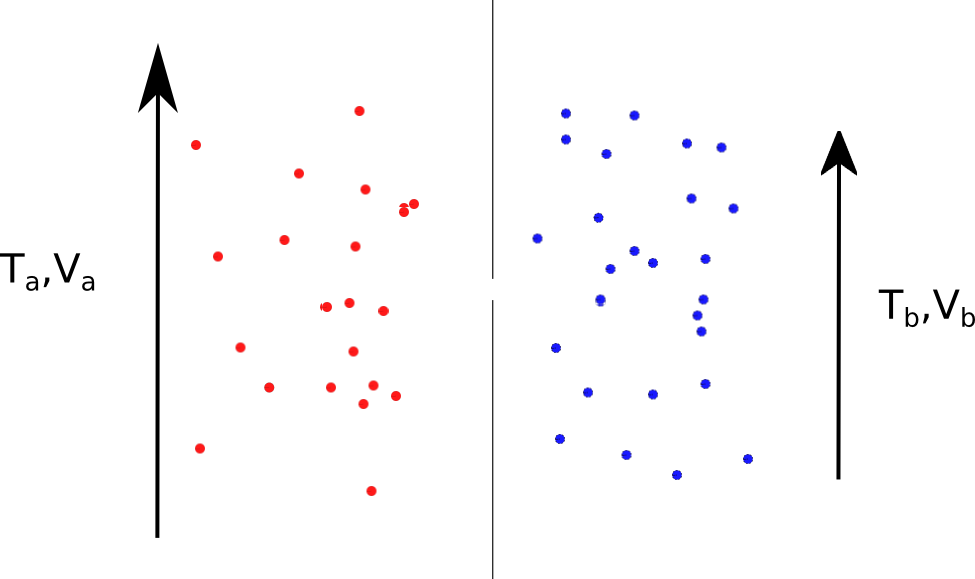}\label{fig1mom}}
\qquad
\subfloat[Colour-coded representation of the macroscopic efficiency for effusion with  momentum transfer for $\eta_C=1/2$. The green dot corresponds to $V_h=5\,\sqrt{{k_B T_h}/{m}}$, $V_c=5.5\,\sqrt{{k_B T_h}/{m}}$ and $\bar{\eta}=0.57$. The stochastic efficiency for the latter case is represented in Fig.~\ref{fig3mom}]{\includegraphics[width=6cm]{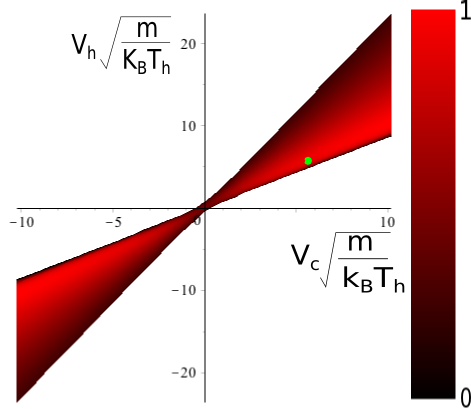}\label{fig2mom}}
\caption{Characteristics of the effusion engine with transversal momentum.}
\end{centering}\end{figure}

The effusion model, discussed in section 2, has also been studied in the presence of  momentum exchange between the reservoirs \cite{wood2007fluctuation},  see Fig.~\ref{fig1mom}.  In this set-up, the gases move with overall average speed $V_h$ and $V_c$ parallel to the separating wall containing the effusion hole. For the purpose of illustration, we assume equal densities and chemical potentials in both reservoirs, and consider a thermal engine, with temperatures $T_h$ and $T_c$ in the respective compartments, driving momentum exchange.

Note that there are $3$ stochastic fluxes: the transfer of particles, of energy and of transversal momentum. This leaves some freedom in the definition of the efficiency. Here, we choose to define it as:
\begin{equation}\eta=-\frac{A_N n+A_{p_x} p_x}{A_U u},\end{equation}
with
\begin{equation}A_N=\frac{3}{2}k_B \ln\left(\frac{T_c}{T_h}\right)+\left(\frac{m V_c^2}{2T_c}-\frac{m V_h^2}{2T_h}\right),\end{equation}
\begin{equation}A_{p_x}=\frac{V_h}{T_h}-\frac{V_c}{T_c},\end{equation}
\begin{equation} A_U=\frac{1}{T_c}-\frac{1}{T_h},\end{equation}
the affinities of the particle, momentum and energy transport, and $n$, $p_x$ and $u$ the amounts of particle, momentum and energy transport. The macroscopic efficiency then reads:
\begin{equation}\bar{\eta}=-\frac{A_N \left\langle n\right\rangle+A_{p_x} \left\langle {p_x}\right\rangle}{A_U \left\langle u\right\rangle},\end{equation}
with $\left\langle n\right\rangle$, $\left\langle p_x\right\rangle$ and $\left\langle u\right\rangle$, the macroscopic particle, energy and momentum fluxes. The total entropy production is given by \cite{wood2007fluctuation}:
\begin{equation}\Delta_i S=A_N \left\langle n\right\rangle+A_{p_x} \left\langle p_x\right\rangle+A_U \left\langle u\right\rangle\geq 0.\end{equation} 
Hence the second law of thermodynamics dictates that macroscopic efficiency in the engine regime is smaller than $1$. This is illustrated for the parameter values considered in  Fig.~\ref{fig2mom}.

\begin{figure}\begin{centering}
\subfloat[Probability distribution $P_t(\eta$) of the efficiency $\eta$ for effusion with momentum transfer.]{\includegraphics[width=6cm]{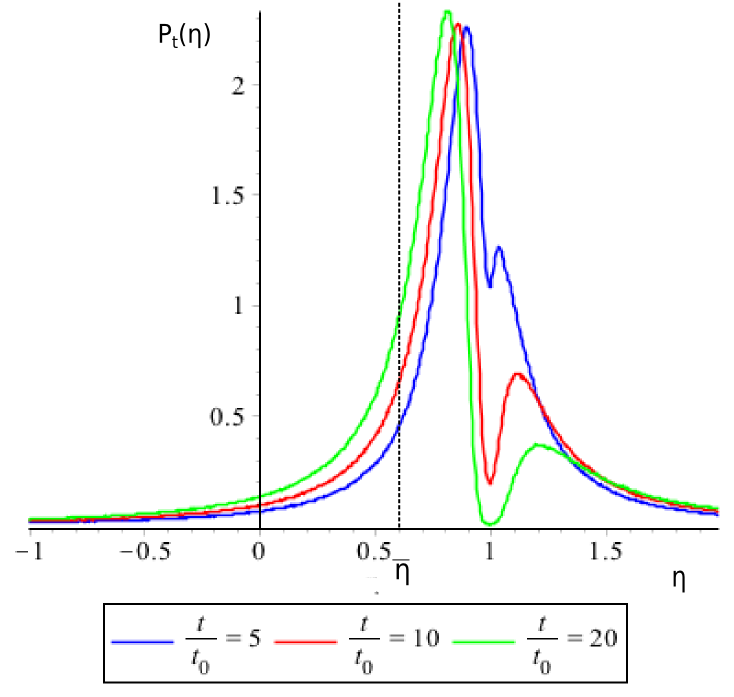}\label{fig3mom}}
\qquad
\subfloat[Approach of $-\ln P_t(\eta)  /t$  for increasing values of $t/t_0$ (blue, red and green curve)  to the large deviation function $J(\eta)$ (black curve). The yellow curve is obtained by extrapolation of the finite time results.]{\includegraphics[width=6cm]{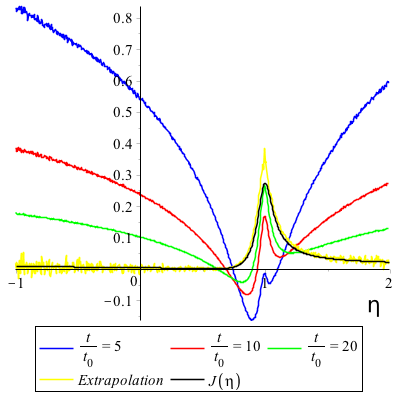}\label{fig4mom}}
\caption{Efficiency fluctuations of effusion with momentum, for $\eta_C=1/2$, $V_h=5\,\sqrt{k_B T_h/m}$ and $V_c=5.5\,\sqrt{k_B T_h/m}$. The macroscopic efficiency is given by $\bar{\eta}=0.57$}
\end{centering}\end{figure}

Following the discussion of the previous examples, we now turn to the efficiency fluctuations. For the finite-time probability distribution $P_t(\eta)$, we again rely  on numerical simulations. The rates of particle transfer are given by:
\begin{eqnarray}T_{h\rightarrow c}(E,p_x)=\frac{1}{t_0}&\frac{\sqrt{2}}{\left(k_B T_h\right)^2\pi \sqrt{m}}\left(E-\frac{p^{2}_x}{2m}\right)^{1/2}\nonumber\\&\exp\left(-\frac{m}{2k_B T_h}\left(\frac{2\left(E-\frac{p^{2}_x}{2m}\right)}{m}+\left(\frac{p_x}{m}+V_h\right)^2\right)\right),\end{eqnarray}
\begin{eqnarray}T_{c\rightarrow h}(E,p_x)=\frac{1}{t_0}&\frac{\sqrt{2}}{\left(k_B T_c\right)^{3/2}\left(k_B T_h\right)^{1/2}\pi \sqrt{m}}\left(E-\frac{p^{2}_x}{2m}\right)^{1/2}\nonumber\\&\exp\left(-\frac{m}{2k_B T_c}\left(\frac{2\left(E-\frac{p^{2}_x}{2m}\right)}{m}+\left(\frac{p_x}{m}+V_c\right)^2\right)\right),\end{eqnarray}
with
$t_0=\sqrt{{2\pi m}/(k_BT_h\sigma^2 \rho^2)}$, the average time between particle crossings from reservoir $h$ to reservoir $c$ and $\rho$ the particle density inside the reservoirs. The results of the simulations are shown in Fig.~\ref{fig3mom}. The minimum at the reversible limit is clearly visible. Note also that, in contrast to effusion without momentum transfer, the probability distribution behaves smoothly around $\eta=0$. The explanation is that, due to the possibility of  momentum transport alone, small efficiencies are possible without net particle transport.

To evaluate the large deviation function of the efficiency, we first note that:
\begin{equation}\left\langle e^{-\lambda_W \Delta W-\lambda_Q \Delta Q}\right\rangle =\left\langle e^{-\lambda_W(A_N\Delta N+A_{p_x}\Delta p_x)-\lambda_q A_U \Delta U}\right\rangle,\end{equation}
and therefore the cumulant generating function of the produced work and heat $\mu_0(\lambda_W,\lambda_Q)$ can be written in terms of the cumulant generating function of the transferred momentum, energy and particle numbers $\mu_1(\lambda_U,\lambda_N,\lambda_{p_x})$:
\begin{eqnarray}\mu_0(\lambda_W,\lambda_Q)&=&\mu_1(A_N \lambda_W,A_{p_x}\lambda_W,A_{U}\lambda_Q)\nonumber\\&=&\sigma\left(\frac{k_B}{2\pi m}\right)^{1/2}\rho T_h^{1/2}\left(1-\frac{G_h\left(A_N \lambda_W,A_{p_x}\lambda_W,A_{U}\lambda_Q\right)}{\left(1+k_BT_hA_{U}\lambda_Q\right)^2}\right)\nonumber\\&&+\sigma\left(\frac{k_B}{2\pi m}\right)^{1/2}\rho T_c^{1/2}\left(1-\frac{G_c\left(A_N \lambda_W,A_{p_x}\lambda_W,A_{U}\lambda_Q\right)}{\left(1-k_BT_cA_{U}\lambda_Q\right)^2}\right),\nonumber\\\end{eqnarray}
with:
\begin{equation}G_h\left(\lambda_N,\lambda_{p_X},\lambda_U\right)=\exp\left(-\lambda_N-\frac{mV_h^2\lambda_U-k_BmT_h\lambda^2_{p_x}+2mV_h\lambda_{p_x}}{2\left(1+k_BT_h\lambda_U\right)}\right),\end{equation}
and
\begin{equation}G_c\left(\lambda_N,\lambda_{p_X},\lambda_U\right)=\exp\left(\lambda_N+\frac{mV_c^2\lambda_U+k_BmT_c\lambda^2_{p_x}+2mV_c\lambda_{p_x}}{2\left(1-k_BT_c\lambda_U\right)}\right).\end{equation}
As was discussed for the effusion model, the large deviation can then be found by numerically contracting the cumulant generating function:
\begin{equation}J(\eta)=-\min_{\lambda}\mu_0(\lambda,\lambda\eta)=-\min_{\lambda}\mu_1(A_N\lambda,A_{p_x}\lambda,A_U\eta\lambda).\end{equation}
The comparison with simulations is shown in  Fig.~\ref{fig4mom}. The extrapolation ansatz seems to work quite well apart from an overshoot at the reversible efficiency.

\section{Mandal-Jarzynski model}
\begin{figure}\begin{centering}
\subfloat[Schematic representation of the Mandal-Jarzynski model with $3$ energy states.]{\includegraphics[width=6cm]{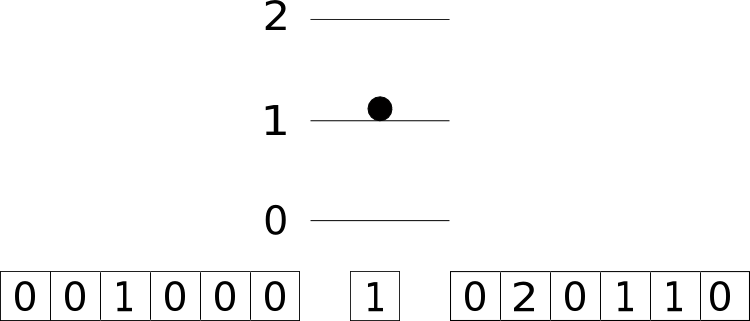}\label{fig1inf}}
\qquad
\subfloat[Alternative physical implementation of the Mandal-Jarzynski model]{\includegraphics[width=6cm]{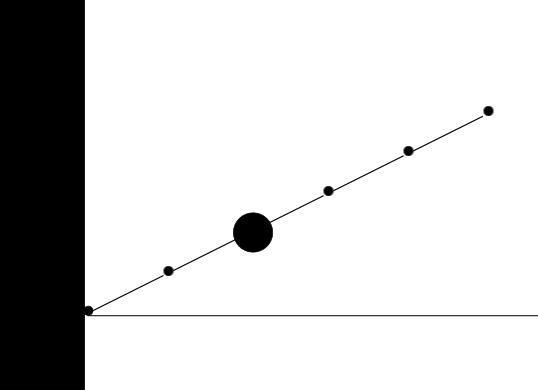}\label{fig1infb}}
\caption{Possible Set-ups for the Mandal-Jarzynski model}
\end{centering}\end{figure}

Recently, there has  been considerable interest in the  stochastic information-to-work conversion. To study this issue in the context of stochastic efficiency, we focus on one of the simplest models, namely the Mandal-Jarzynski engine \cite{mandal2012work}, cf.~Fig.~\ref{fig1inf}. 

A particle, in contact with a heat bath at constant temperature $T$, can make transitions between a number of energy levels, which are multiples of $\delta E$.  Apart from the thermal dissipative transitions due to the bath, the particle undergoes transitions that are driven by the entries of a linear tape. This interaction corresponds to a form of input work. 
The energy levels of the systems are numbered as shown in Fig.~\ref{fig1inf}. The input tape consists of a sequence of entry values, each referring uniquely to one of the possible energies, which are presented subsequently to the system. During such an interaction, the particle  is moved to the energy level with number equal to the initial value on the tape. In this way, work is delivered. After each entry of the tape, the system is left to thermalize, and the new energy state of the particle is written into the tape, which we call the final or exit value. Note that this will alter the Shannon entropy of the tape. The tape moves on one step and the process is repeated.   We mention a closely related implementation of the Mandal-Jarzynski model, shown in Fig.~\ref{fig1infb}. A Brownian particle in contact with a thermal bath at temperature $T$, can make jumps on a discrete lattice with a wall at position zero and a force gradient in the negative direction, inducing an energy difference between two neighbouring sites  equal to $\delta E$. After the particle has thermalized, its position is measured, and the wall is instantaneously brought to the measured position. Next the wall is adiabatically moved back to the original position with delivery of work,  after which the process is repeated. While we studied both versions, we focus in the following on the simplest situation giving nontrivial results for stochastic efficiency, namely a Mandal-Jarzynski model with 3 states, cf.~Fig.~\ref{fig1inf}.

It is clear that the amount of delivered work after one entry (trit) of the tape is given by:
 \begin{eqnarray} w=\delta E\left(f-i\right),\end{eqnarray}
 where $i$ and $f$ are the initial and final state of the system, respectively.  The corresponding information entropy production in the tape can be defined in multiple ways, but if we assume that the dynamics is the same for the reversed trajectory, it is given by \cite{barato2014unifying,horowitz2013imitating}:
 \begin{equation}\Delta s_{tape}=k_B\left(\ln\left(p_{I,i}\right)-\ln\left(p_{I,f}\right)\right),\end{equation}
with $i$ and  $f$ the initial and final state of the trit and  $p_{I,k}$ the probability for entry state $k$ on the tape. The average information entropy change upon processing one trit is thus given by:
\begin{eqnarray}\left\langle\Delta s_{tape}\right\rangle&=&k_B\sum_j \left(p_{I,j}-p_{F,j}\right)\ln\left(p_{I,j}\right).\end{eqnarray}
Here:
 \begin{equation}
p_{F,j}=e^{-j\frac{\delta E}{k_B T}}/\left(1+e^{-\frac{\delta E}{k_B T}}+e^{-2\frac{\delta E}{k_B T}}\right),
\end{equation}
is the probability that the trit leaves the system in state j ($j=0,1,2$). The information entropy written in this way is the sum of the change in Shannon entropy and the entropy production of an auxiliary Mandal-Jarzynski system if it were to bring the tape back to its original distribution \cite{barato2014stochastic}.

The efficiency is defined as the ratio of the delivered work and the amount of information consumed.
\begin{figure}\begin{centering}
\includegraphics[width=6cm]{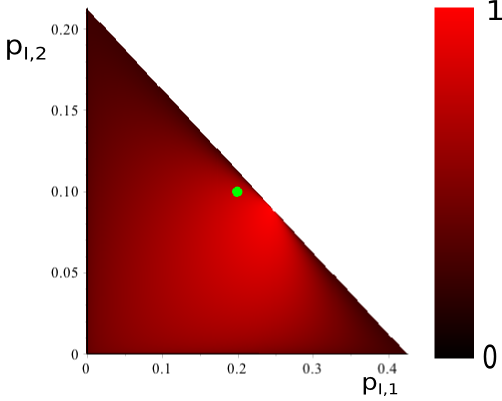}
\caption{Macroscopic efficiency of the Mandal-Jarzynski model in the engine regime, with $\delta E=k_B T$. The green dot corresponds to $p_{I,1}=0.2$, $p_{I,2}=0.1$ and $\bar{\eta}=0.68$. The stochastic efficiency for the latter case is represented in Fig.~\ref{fig3mom}}
\label{fig2inf}
\end{centering}\end{figure}
In particular the macroscopic efficiency is given by:
\begin{eqnarray}\bar{\eta}&=&\frac{\left\langle w\right\rangle}{T\left\langle\Delta s_{tape}\right\rangle}\nonumber\\&=&\frac{\delta E}{k_B T}\frac{\left(2\left(p_{F,2}-p_{I,2}\right)+\left(p_{F,1}-p_{I,1}\right)\right)}{\left(\left(p_{I,0}-p_{F,0}\right)\ln\left(p_{I,0}\right)+\left(p_{I,1}-p_{F,1}\right)\ln\left(p_{I,1}\right)+\left(p_{I,2}-p_{F,2}\right)\ln\left(p_{I,2}\right)\right)}.\nonumber\\\end{eqnarray}
A colour-coded plot is given in Fig.~\ref{fig2inf}, for $\delta E/k_B T=1$. Reversible efficiency ($\bar{\eta}=1$) can be reached in the limit where $p_{I,j}=p_{F,j}$, $j=1,2$.

\begin{figure}\begin{centering}
\subfloat[Probability distribution of the efficiency.]{\includegraphics[width=6cm]{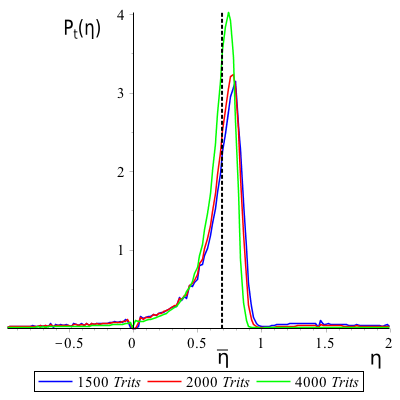}\label{fig4inf}}
\qquad
\subfloat[Approach of $-\ln P_t(\eta)  /t$  for increasing values of $t/t_0$ (blue, red and green curve)  to the large deviation function $J(\eta)$ (black curve). The yellow curve is obtained by extrapolating the finite time results.]{\includegraphics[width=6cm]{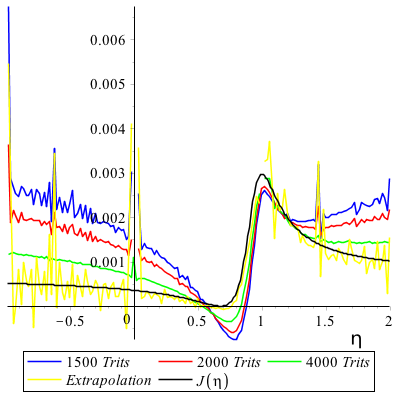}\label{fig5inf}}
\caption{Efficiency fluctuations of the Mandal-Jarzynski model, for $E=k_BT$,$p_{I,1}=0.2$ and $p_{I,2}=0.1$ . The macroscopic efficiency is given by $\bar{\eta}=0.68$}
\end{centering}\end{figure}
The joint probability distribution $P_{Init,N}\left(n_{1},n_{2}\right)$ of $n_{1}$ and $n_{2}$ incoming trits with value 1 and 2 respectively on a total of N trits is given by:
\begin{equation}\label{PInit}P_{Init,N}(n_1, n_2)=\frac{N!}{(N-n_1-n_2)!n_1!n_2!}p_{I,1}^{n_1}p_{I,2}^{n_2}(1-p_{I,1}-p_{I,2})^{N-n_1-n_2}\end{equation}
and the joint probability distribution of $n_{1}$ and $n_{2}$ outgoing trits with value 1 and 2 respectively on a total of N trits is given by:
\begin{eqnarray}P_{Final,N}(n_1, n_2)&=&\frac{N!}{(N-n_1-n_2)!n_1!n_2!}p_{F,1}^{n_1}p_{F,2}^{n_2}(1-p_{F,1}-p_{F,2})^{N-n_1-n_2},\nonumber\\\end{eqnarray}
which is independent of the distribution of the incoming trits. Using these distributions, numerical simulations can be performed, to evaluate the probability distribution of the stochastic efficiency $\eta=w/(T \Delta s_{tape})$, cf.~Fig.~\ref{fig4inf}. The minimum around $\eta=0$ is analogous to the minimum in the effusion model, and disappears in the large time limit. Furthermore, one observes the weak local minimum at reversible efficiency $1$ with, to its right, a more clearly visible maximum. 

The large deviation functions of $x_1=n_{I,1}/N$ and $x_2=n_{I,2}/N$ incoming trits with value 1 and 2 respectively, is found from equation (\ref{PInit}):
\begin{eqnarray}\gamma_{Init}(x_1,x_2)=x_1 && \ln(x_1)+x_2 \ln(x_2)+(1-x_1-x_2) \ln(1-x_1-x_2)\nonumber\\&&-x_1 \ln(p_{I,1}) -x_2\ln(p_{I,2})-(1-x_1-x_2)\ln(1-p_{I,1}-p_{I,2}),\nonumber\\\end{eqnarray}
and a completely analogous expression for the large deviation function $\gamma_{Final}(y_1,y_2)$ of $y_1=n_{F,1}/N$ and $y_2=n_{F,2}/N$, with $p_{I,k}$ replaced by $p_{F,k}$. From this, the large deviation function of the efficiency can be calculated:
\begin{equation}J(\eta)=\min_{x_1,x_2,y_1,y_2}\left(\gamma_{Init}(x_1,x_2)+\gamma_{Final}(y_1,y_2)\right)\end{equation}
where $x_1$, $x_2$, $y_1$ and $y_2$ are constrained to reproduce the efficiency $\eta$ under consideration, i.e., they are related by:
\begin{equation} \eta=\frac{\delta E}{k_B T}\frac{\left(2\left(y_2-x_2\right)+y_1-x_1\right)}{\left(x_0-y_0\right)\ln\left(p_{I,0}\right)+\left(x_1-y_1\right)\ln\left(p_{I,1}\right)+\left(x_2-y_2\right)\ln\left(p_{I,2}\right)}.\end{equation}
This minimisation can be done numerically, cf.~Fig.~\ref{fig5inf}. In spite of the noise, which is mainly due to the discreteness of variables, the extrapolation seems to work quite well. Also, the telltale maximum in the large deviation function close to reversible efficiency is again reproduced.

\section{Discussion}
The concept of efficiency plays a crucial role in thermodynamics, especially when the efficiency is defined in such a way that is leads to universal system-independent statements, such as the one concerning Carnot efficiency. With the advent of stochastic thermodynamics, it  is natural to revisit such questions for stochastic efficiency. Universal statements appear to be possible for the large deviation function characterising the asymptotic time regime. In particular, long-time realisations with reversible efficiency are exponentially least probable. One purpose of this paper has been to verify and document the salient features of the time-asymptotic stochastic efficiency in five different settings, namely driven Brownian motion, effusion with a thermo-chemical and thermo-velocity gradient, a quantum dot and a model for information to work conversion. In addition, we provide the analysis for finite time including the approach to and extrapolation into the asymptotic time regime. A revealing feature of our analysis is that the large deviation properties can be obtained quite consistently by extrapolation from rather short finite time results.  The other encouraging message is that one can apply the analysis to a wide variety of completely different implementations, some of which may be easier to realise. Both observations imply that an experimental verification should not pose a real problem. In particular, in view of existing experiments on the issue \cite{berut2012experimental,Toyabe,wang,seifert,dellago}, the experimental implementation for the stochastic efficiency of an information to work engine should be relatively straigthforward.

\ack
We thank Thijs Becker and Bart Cleuren for a careful reading of the manuscript.

\setcounter{section}{1}
\appendix

\section{Large deviation functions from finite time results\label{AppLDF}}
The central object of interest in stochastic efficiency is its large deviation function. Unfortunately, the evaluation of exponentially unlikely events is obviously very difficult. 
For this reason, we propose a simple and apparently robust method to deduce the large deviation function $J(\eta)$ from finite time results. As input we give the efficiency probability distributions for three finite times $t_1$, $t_2$ and $t_3$. We propose the following ansatz for  $P_t(\eta)$ :
\begin{equation}P_{\mathrm{Fit},t}(\eta)=A(\eta)t^{-B(\eta)}e^{-tJ(\eta)}\end{equation}
Here, $A(\eta)$, $B(\eta)$ and $J(\eta)$ are three fitting parameters for each $\eta$. $J(\eta)$ shall be our estimate for the large deviation function. The fitting parameters can 
be obtained from  the known values $-\ln\left(P_{t_i}(\eta)\right)$, with $i=1,2,3$,  since:
\begin{equation}\left[
\begin{array}{c}
-\ln\left(P_{t_1}(\eta)\right)/t_1\\
-\ln\left(P_{t_2}(\eta)\right)/t_2\\
-\ln\left(P_{t_3}(\eta)\right)/t_3
\end{array}
\right]
=
\left[
\begin{array}{ccc}
1&1/t_{1}&\ln  \left( t_{1} \right)/t_{1}\\
1&1/t_{2}&\ln  \left( t_{2} \right)/t_{2}\\
1&1/t_{3}&\ln  \left( t_{3} \right)/t_{3}\\
\end {array}\right]
\left[\begin{array}{c}J(\eta)\\-\ln\left(A(\eta)\right)\\B(\eta)\end{array}\right].
\end{equation}
Inverting this matrix equality leads to an estimate of the large deviation function.

\section{Efficiency calculations from event probability distributions \label{AppEve}}
Consider a model consisting of $k$ tight-coupled processes (e.g. k channels for particle transport), where the $i$-th process undergoes $n_i$ events, $i=1,..,k$. Furthermore, the total delivered amount of work and heat can be written as $W(n_1,...,n_k)$ and $Q(n_1,...,n_k)$. Once the event probability distribution is known, the probability distribution of $\eta$ can be written as:
\begin{equation}\label{PetaGen2}P_t(\eta)=\sum_{n_1,...,n_k}P_t(n_1,...,n_k)\delta_{\eta Q(n_1,...,n_k),W(n_1,...,n_k)},\end{equation}
for discrete variables and
\begin{equation}\label{PetaGen}P_t(\eta)=\int dn_1..dn_kP_t(n_1,...,n_k)\delta\left(\eta-\frac{W(n_1,...,n_k)}{Q(n_1,...,n_k)}\right),\end{equation}
for continuous variables, where $P_t(n_1,...,n_k)$ is the probability that at time t, for process i, $n_i$ events have occured.
Using the corresponding event large deviation function $\gamma (n_1,...,n_k)=-\lim_{t\rightarrow\infty}\left(1/t \ln\left(P_t(n_1,...,n_k)\right)\right)$, the efficiency large deviation function $J(\eta)$ can then be calculated via the contraction principle:
\begin{equation}J(\eta)=\min_{n_1,...,n_k}\gamma(n_1,...,n_k),\label{JetaApp}\end{equation}
where $n_1,...,n_k$ are conditioned to:
\begin{equation}\eta=\frac{W(n_1,...,n_k)}{Q(n_1,...,n_k)}.\end{equation}
This provides a good scheme for numerical calculations. 

We proceed to  show that there are, in this case, only two extrema of the efficiency large deviation function: the macroscopic efficiency and the reversible efficiency. 
Using Lagrange multipliers, we have:
\begin{equation}J(\eta)=\mathrm{Extr}_{x_1,...,x_k,\lambda}L(x_1,..,x_k,\lambda),\end{equation}
with
\begin{equation}L(x_1,...,x_k,\lambda)=\gamma(x_1,...,x_k)+\lambda\left(\eta Q(x_1,...,x_k)-W(x_1,...,x_k)\right).\end{equation}
Therefore, $x_1,...,x_k,\lambda$ are constrained to:
\begin{equation}\frac{\partial \gamma(x_1,...,x_k)}{\partial x_i}=-\lambda\left(\eta \frac{\partial Q(x_1,...,x_k)}{\partial x_i}-\frac{W(x_1,...,x_k)}{\partial x_i}\right),\label{GamApp}\end{equation}
for all i. As $L(x_1,...,x_k,\lambda)$ is an extremum of $\lambda$ and $x_i$, $i=1,...,k$, we have
\begin{eqnarray}\frac{d}{d\eta}J(\eta)&=&\sum_i\frac{\partial L(x_1,...,x_k,\lambda)}{\partial x_i}\frac{\partial x_i}{\partial \eta}+\frac{\partial L(x_1,...,x_k,\lambda)}{\partial \lambda}\frac{\partial \lambda}{\partial \eta}+\frac{\partial L(x_1,...,x_k,\lambda)}{\partial \eta}\nonumber\\&=&\frac{\partial L(x_1,...,x_k,\lambda)}{\partial \eta}\nonumber\\&=&\lambda Q(x_1,...,x_k).\end{eqnarray}
Note that $Q(x_1,...,x_k)=0$ corresponds to reversible efficiency due to the fluctuation theorem, and that $\lambda=0$ corresponds to $\partial \gamma(x_1,...,x_k)/\partial x_i=0$ (using equation \ref{GamApp}) which is equivalent with macroscopic efficiency (as we assume convex event large deviation functions). As one of these two equalities has to be fulfilled to be in an extremum of $J(\eta)$, we conclude that these are the only two extremums of the large deviation function

We finally note that the expression for $J(\eta)$ can be further simplified if the processes are independent. We shall illustrate this for systems consisting of two independent processes (with $n_1$ and $n_2$ events respectively). The extension to more independent processes is straightforward. The amount of delivered work per event in the $i$th process is written as $w_i$ and the amount of extracted heat is $q_i$, $i=1,2$. The efficiency is then given by:
\begin{eqnarray}\eta&=&\frac{w_1 n_1+w_2 n_2}{q_1 n_1+q_2 n_2}\nonumber\\&=& \frac{q_1 n_1}{q_1 n_1+q_2 n_2} \eta_1+n_2 \frac{q_2 n_2}{q_1 n_1+q_2 n_2} \eta_2\nonumber\\&=& \alpha\eta_1+(1-\alpha) \eta_2,\end{eqnarray}
with
\begin{equation}\label{Fora}\alpha=\frac{q_1 n_1}{q_1 n_1+q_2 n_2}=\frac{\eta-\eta_2}{\eta_1-\eta_2},\end{equation}
and $\eta_i=w_i/q_i$, $i=1,2$.
Therefore, the probability distribution of $\eta$ at time t is given by:
\begin{eqnarray}P_t(\eta)&=&\int\int dw dq P_t(w,q)\delta\left(\eta-\frac{w}{q}\right)\nonumber\\&=&\int\int dn_1 dn_2\left|w_1 q_2-w_2 q_1\right|P_t (n_1)P_t (n_2) \delta\left(\eta-\alpha\eta_1-(1-\alpha) \eta_2\right).\nonumber\\ \end{eqnarray}
From large deviation theory, we can now write the efficiency large deviation function $J(\eta)$ in terms of terms of the large deviation functions of the event numbers $\varphi_1(n_1)$ and $\varphi_2( n_2)$:
\begin{equation} J(\eta)=\min_x \left(\varphi_1\left(\frac{\alpha x}{ q_1}\right)+\varphi_2\left(\frac{(1-\alpha)x}{q_2}\right)\right).\end{equation}

\section{Event large deviation function of quantum dot-like models\label{AppQD}}
We present for completeness the large deviation function for the number of particles travelling through one energy level of a quantum dot (with energy $E$ and coupling constant $\Gamma=1$) . We  only quote the final result, as similar calculations can be found in the literature, see e.g. \cite{willaert2014fluctuation,lacoste2008fluctuation,harris2007fluctuation,verley2013modulated}. We recall  that the fluxes are given by equations (\ref{RateQD1}) and (\ref{RateQD2}). The cumulant generating function is given by:
\begin{equation}f(\gamma)=k+\sqrt{r+ q(\gamma)s},\end{equation}
with
\begin{eqnarray} k=-\frac{\left(k^{+}_L+k^{-}_L+k^{+}_R+k^{-}_R\right)}{2};\quad s=\sqrt{k^{+}_Lk^{+}_Rk^{-}_Lk^{-}_R};\quad \gamma_0=\ln\left(\frac{k^+_Lk^-_R}{k^+_Rk^-_L}\right)\nonumber\\\rho(\gamma)=e^{\gamma_0/2-\gamma};\quad r=k^2-k^+_Lk^-_R-k^+_Rk^-_L;\quad q(\gamma)=\rho(\gamma)+\rho(\gamma)^{-1}.
\end{eqnarray}
The large deviation function reads:
\begin{equation}h(n)=-f(\gamma(n))-n\gamma(n),\end{equation}
with:
\begin{equation}\bar{q}(n)=\frac{2n}{s}\sqrt{r+2n^2+2\sqrt{s^2+r n^2+n^4}},\end{equation}
\begin{equation}\gamma(n)=\frac{\gamma_0}{2}-\ln\left(\frac{\sqrt{\bar{q}(n)^2+4}+\bar{q}(n)}{2}\right)\end{equation}

\end{document}